\input harvmac

\def\third{{\textstyle{1\over3}}}

\def\bfone{\relax{\rm 1\kern-.35em 1}}
\def\inbar{\vrule height1.5ex width.4pt depth0pt}
\def\IC{\relax\,\hbox{$\inbar\kern-.3em{\mss C}$}}
\def\ID{\relax{\rm I\kern-.18em D}}
\def\IF{\relax{\rm I\kern-.18em F}}
\def\IH{\relax{\rm I\kern-.18em H}}
\def\II{\relax{\rm I\kern-.17em I}}
\def\IN{\relax{\rm I\kern-.18em N}}
\def\IP{\relax{\rm I\kern-.18em P}}
\def\IQ{\relax\,\hbox{$\inbar\kern-.3em{\rm Q}$}}
\def\IR{\relax{\rm I\kern-.18em R}}
\font\cmss=cmss10 \font\cmsss=cmss10 at 7pt
\def\ZZ{\relax\ifmmode\mathchoice
{\hbox{\cmss Z\kern-.4em Z}}{\hbox{\cmss Z\kern-.4em Z}}
{\lower.9pt\hbox{\cmsss Z\kern-.4em Z}}
{\lower1.2pt\hbox{\cmsss Z\kern-.4em Z}}\else{\cmss Z\kern-.4em
Z}\fi}

\Title{\vbox{\baselineskip12pt
\hbox{CERN-TH/97-157}
\hbox{RU-97-4-B}
\hbox{hep-th/9707125}
}}
{\vbox{\centerline{$U$-branes and $T^3$ fibrations}}}

\centerline{James T. Liu$^a$ and Ruben Minasian$^{b}$}

\bigskip\centerline{$^a${\it The Rockefeller University, New York, NY
10021, USA}}
\medskip\centerline{$^b${\it CERN, CH-1211 Geneva 23, Switzerland}}

\vskip .3in

We describe eight-dimensional vacuum configurations with varying moduli
consistent with the $U$-duality group $SL(2,\ZZ) \times SL(3,\ZZ)$. Focusing
on the latter less-well understood $SL(3,\ZZ)$ properties, we construct a
class of fivebrane solutions living on lines on a three-dimensional base
space.  The resulting $U$-manifolds, with five scalars transforming under
$SL(3)$, admit a Ricci-flat K\"ahler metric.  Based on the connection with
special lagrangian $T^3$ fibered Calabi-Yau 3-folds, this construction
provides a simple framework for the investigation of Calabi-Yau mirrors.

\vskip .3in
\Date{\vbox{\baselineskip12pt
\hbox{CERN-TH/97-157}
\hbox{July 1997}}}

\def\sqr#1#2{{\vbox{\hrule height.#2pt\hbox{\vrule width
.#2pt height#1pt \kern#1pt\vrule width.#2pt}\hrule height.#2pt}}}

\lref\witdiv{E. Witten, {\sl String Theory Dynamics in Various
Dimensions}, Nucl. Phys. B 443 (1995) 85.}

\lref\htu{C.M. Hull and  P.K. Townsend,  {\sl Unity of Superstring
Dualities}, Nucl. Phys. B 438 (1995) 109.}

\lref\gsvy{B.R. Greene, A. Shapere, C. Vafa and S.T. Yau, {\sl Stringy
Cosmic String and Noncompact Calabi-Yau Manifolds}, Nucl. Phys. B 337
(1990) 1.}

\lref\vafaff{C. Vafa, {\sl Evidence for F theory}, Nucl. Phys. B 469
(1996) 403.}

\lref\kv{A. Kumar and C. Vafa, {\sl U-Manifolds}, Phys. Lett. B 396 (1997)
85.}

\lref\bbs{ K. Becker, M. Becker and A. Strominger, {\sl Five-branes,
Membranes and Nonperturbative String Theory}, Nucl. Phys. B 456 (1995)
130.}

\lref\syz{A. Strominger, S.T. Yau and E.  Zaslow, {\sl Mirror Symmetry is
T duality}, Nucl. Phys. B 479 (1996) 243.}

\lref\gw{M. Gross and P.M.H. Wilson, {\sl  Mirror Symmetry via 3-tori for
a Class of Calabi-Yau Threefolds}, {\tt alg-geom/9608004.} }

\lref\morr{D.R. Morrison, {\sl The Geometry Underlying Mirror Symmetry},
{\tt alg-geom/9608006.} }

\lref\solv{L. Andrianopoli, R. D'Auria, S. Ferrara, P. Fre and M.
Trigiante, {\sl R-R Scalars, U Duality and Solvable Lie Algebras},
Nucl. Phys. B 496 (1997) 617.}

\lref\solvv{L. Andrianopoli, R. D'Auria, S. Ferrara, P. Fre, R. Minasian
and M. Trigiante, {\sl Solvable Lie Algebras in Type IIA, IIB and M
Theories}, Nucl. Phys. B 493 (1997) 249.}

\lref\bho{E. Bergshoeff, C. Hull and T. Ortin, {\sl Duality in the Type II
Superstring Effective Action}, Nucl. Phys. B 451 (1995) 547.}

\lref\grassi{A. Grassi, {\sl On Minimal Models of Elliptic Threefolds},
Math. Ann. 290 (1991) 287.}


\newsec{Introduction}

$U$-duality has played an important role for understanding nonperturbative
aspects of string theory \refs{\htu, \witdiv}. This role is not just
restricted to establishing the relationships between different theories, but
also to constructing new vacua in string theory. The most dramatic example of
this is $F$-theory \vafaff, which is based on the $SL(2,\ZZ)$ $U$-duality
group of type $IIB$ theory. The dilaton is allowed to jump according to
$SL(2,\ZZ)$, and the presence of seven-branes in the theory (which are
direct generalizations of so called stringy cosmic strings \gsvy) ensures
the existence of well-defined vacua. The jumps are encoded in a manifold
that has a fibered structure where the fiber is the geometrization of the
duality group. It has been shown recently \kv\ that this type of argument
can be applied not just to the type $IIB$ theory, but to other theories as
well, and in particular to $N=2$ theories in eight and seven dimensions.
The structure of the moduli spaces, given by $SL(2,\ZZ)\times SL(3,\ZZ)
\backslash SL(2) \times SL(3)/SO(2) \times SO(3)$ and $SL(5,\ZZ)
\backslash SL(5)/SO(5)$ respectively, leads to arguments in favor of new
higher-dimensional theories formulated on $U$-manifolds admitting
$T^2 \times T^3$ and $T^5$ fibers.
Putting aside the practical questions on construction of new vacua, in the
following we try to understand the relation between supersymmetry and the
fibration structure of the resulting $U$-manifolds.

We concentrate on the eight-dimensional case with the duality group
$SL(2,\ZZ) \times SL(3,\ZZ)$. While the first factor is well-understood,
there is not much known about the solutions that respect the second. The
five-dimensional piece of the moduli space parametrizes a three-torus at
constant volume;
that is, a supersymmetric three-cycle \refs{\bbs, \syz} in the resulting
$U$-threefold (see also \refs{\gw, \morr}). As a first step in understanding
this, we construct a family of fivebrane solutions that transform
consistently with $SL(3,\ZZ)$. This family lives on a three-dimensional base
that is topologically $S^3$, but each individual member is of real
codimension two on the base (in agreement with the naive expectation
that a fivebrane solution in eight dimensions should depend on only two
transverse coordinates).  In particular, the solutions all take the form
of overlapping fivebranes, each individually preserving half of the
supersymmetries, but together preserving only a quarter.  This is in
agreement with the expectation that intersecting branes break additional
supersymmetries.  As a result, this leaves us with a $D=3$ $N=2$ theory
upon compactification on a $S^2\times S^3$ base consistent with a full
$SL(2,\ZZ)\times SL(3,\ZZ)$ solution.

While the overlapping fivebranes are constructed based on first order
supersymmetry equations, it turns out that these exact same equations are
in perfect correspondence with the conditions for the resulting $U$-manifold
to admit a special lagrangian $T^3$ fibration.  This should in fact come as
no surprise, as $SL(3)/SO(3)$ is (at least locally) the moduli space of
$T^3$.  Based on this connection with $T^3$ fibered Calabi-Yau 3-folds, we
give an explicit realization of mirror symmetry based on $T$-duality on the
fibers.  This picture is particularly nice, as it is manifest how the
complex structure and K\"ahler deformations are interchanged in the mirror
pairs.

In the next section we give an overview of $U$-scalars and the role of
solvable Lie algebras in their classification.  Then in section 3 we
specialize to eight dimensions and the construction of the $SL(3,\ZZ)$
based overlapping fivebrane solutions.  This discussion is simplified by the
use of first order Killing spinor equations.  At this point we also indicate
the straightforward $SL(n, \ZZ)$ generalization.  In section 4 we connect
the $U$-manifold discussion with $T^n$ fibered Calabi-Yau manifolds by
demonstrating the correspondence of the first order equations with the
special lagrangian conditions.  This then enables us to consider the action
of $T$-duality in generating mirror pairs.  Finally, we conclude with some
comments on global issues that are still not completely understood.

\newsec{$U$-duality and scalar manifolds}

We consider supersymmetric theories whose moduli spaces are given by
homogeneous coset manifolds $U_D/H_D$, where $U_D$ is a non-compact Lie
group and $H_D$ is a maximal compact subgroup. The moduli spaces in
these theories are exact since supersymmetry protects again quantum
corrections. The properties of the moduli space have been used to
construct a large class of new vacua, in particular by introducing
($D-3$) branes carrying scalar charges. It has also been shown that
they realize compact fibered manifolds, where the fiber captures the
symmetries of the moduli space.

Our main focus is on maximal supersymmetric theories in eight and seven
dimensions.  Restricting to cosets of the type $SL(n)/SO(n)$, it
is easy to notice that the scalar manifold $\cal M$ can be seen as
originating from an ``internal'' $T^n$ torus.
Indeed,  when the other fields are set to zero (note that these should
respect $U$ since it is the symmetry of the full theory), all
the solutions can be understood from pure gravity in $n$ dimensions
higher, provided that the volume of the torus is kept fixed. As it
turns out, these internal tori can always be realized as special lagrangian
submanifolds of a larger $U$-manifold, and extended supersymmetric
theories
provide a natural environment for these. Although the construction
developed in this paper is best suited for the case when the dimension of
the cycle is half that of the manifold, this does not need to
be the case --- numerous examples of elliptically fibered $n$-folds
have been discussed in $F$-theory literature. One may also consider
$T^3$ fibrations in the manifolds of $G_2$ holonomy, but these will not
be addressed here.

The concept of solvable algebras $Solv(U/H)$ that associate the
scalars of the coset to group generators \refs{\solv, \solvv} has proven to be
useful in analyzing the properties of $U$-scalars.
This identification of coset manifolds with the group manifolds of
solvable Lie algebras leads to
replacing some of the notion of geometry of cosets by algebraic ones.  In
particular, this allows one to count the precise number of translational
symmetries. As a matter of fact, the translational symmetries of $U_D/H_D$
in $D$ dimensions are classified by $U_{D+1}$, and, somewhat
surprisingly, one finds that only half of the $RR$ scalars have them (the
$NSNS$ vs. $RR$ division is classified by $O(9-D,9-D)$). The scalar manifolds
appearing in the toroidal compactifications with maximal supersymmetry are
non-compact homogeneous spaces of maximal rank $11-D$, and the associated
solvable algebras have a special structure. It is of special importance that
the type $IIB$ theory in ten dimensions and $N=2$ theories in eight and
seven have these two special features. Even though most of our results
are obtained explicitly for $SL(3,\ZZ)$, the generalization to the
$SL(5,\ZZ)$ case is straightforward.

Specializing to the $SL(3,\ZZ)$ case, one would like to understand
the structure of the moduli space, and in particular its mapping on the
base for the fivebranes.  The study of the $SL(2,\ZZ)$ case
\gsvy\ reveals the special importance of the orbifold points and the
uncontractable cycles on the moduli space. Much remains to be done in
the study of global properties of the $SL(3,\ZZ)$ case. But by focusing
on local issues below, we nevertheless gain at least a partial understanding
of the fibration structure of the $U$-manifold, and furthermore develop new
insight on Calabi-Yau mirror symmetry.

\newsec{Fivebranes on $S^3$}

As a starting point for constructing the fivebrane solutions, we begin with
a description of the effective action for the scalar fields.  While the
eight-dimensional $N=2$ theory contains a total of seven scalars, we only
consider the explicit action for the five scalars corresponding to the
$SL(3,\IR)/SO(3)$ coset.  These five scalars may be represented in terms of
a vielbein, $V_{ai}$, with determinant 1.  Here $a$ is a $SO(3)$ index,
while $i$ is a $SL(3)$ index.  Since $V$ is essentially a coset
representative, we define
\eqn\mauer{
(\partial_\mu^{\vphantom{()}} VV^{-1})^{ab} = P_\mu^{(ab)} + Q_\mu^{[ab]},
}
where $P_\mu$ is symmetric in the $SO(3)$ indices, and is used in
constructing the kinetic term for the scalars.  $Q_\mu$ is antisymmetric and
is a composite $SO(3)$ connection.  Due to the $SO(3)$ invariance, it is
clear that $V$ contains only five scalar degrees of freedom.

In terms of the vielbein $V$, the eight-dimensional effective action for
the scalar fields coupled to gravity may then be written as
\eqn\action{{\cal L} = {1\over2\kappa^2}\sqrt{-g}[R-\Tr P_\mu P^\mu+\cdots],}
and gives rise to the following equations of motion:
\eqn\eom{\eqalign{{\cal D}^\mu P_\mu &= 0\cr
R_{\mu\nu}&=\Tr P_\mu P_\nu,}}
where ${\cal D}$ is the Lorentz and $SO(3)$ covariant derivative, so that
\eqn\dls{
{\cal D}_\mu P_\nu= \nabla_\mu P_\nu+ [Q_\mu,P_\nu].}

In principle, we are interested in fivebrane solutions that solve the above
equations of motion.  However, in contrast with the $SU(1,1)/U(1)$ $F$-theory
with a complex moduli space, since the $SL(3,\IR)/SO(3)$ space is
odd-dimensional, complex geometry no longer plays a dominant role in
constructing the solutions.  This is a rather crucial difference between
the $SL(3,\ZZ)$ $U$-dual solutions and the $SL(2,\ZZ)$ solutions, and forces
us to develop new methods in the present case.

With five scalars coupled to gravity, the second order equations of motion
are rather cumbersome to examine.  Fortunately we may use supersymmetry as a
guide, and examine the first order Killing spinor equations.  Recall that
a single fivebrane solution may be constructed based on preserving exactly
half of the eight-dimensional supersymmetries.  Furthermore, by now many
cases of overlapping branes are well understood from a supersymmetric
point of view.  The rest of this section details the construction of
overlapping fivebranes, and their relation to half-supersymmetry projections.

The supersymmetry of the theory given by \action\ may be obtained in several
manners.  A direct compactification of type $IIB$ theory to eight dimensions
will give an explicit parametrization of the vielbein $V$ along with its
supersymmetry properties.  For completeness this reduction is presented in
the Appendix.  On the other hand, making use of T duality, we may equally
well relate the action to $M$-theory compactified on a 3-torus, in which case
the $SL(3,\IR)/SO(3)$ coset is directly related to the symmetries of the
compactification $T^3$.  We will have more to say about this later.  In
either case, the eight-dimensional fermions are pseudo-Majorana, and
transform as a doublet under $SO(3)$.  Using $T^a$ to denote representation
matrices for the spinor representation of $SO(3)$, the resulting supersymmetry
transformations on the fermions are given by
\eqn\susy{\eqalign{\delta\chi^a &= -\half\gamma^\mu
P_\mu^{ab}T^b\epsilon\cr
\delta\psi_\mu^{\vphantom{ab}} &= {\cal D}_\mu^{\vphantom{ab}} \epsilon
\equiv [\nabla_\mu^{\vphantom{ab}} + {1\over4}Q_\mu^{ab}T^{ab}]\epsilon.}}
Note that the spin-1/2 fermions, $\chi^a$, carry an additional vector index
$a$ of $SO(3)$.

A basic fivebrane solution, preserving exactly half of the supersymmetries,
may be obtained by demanding the vanishing of $\delta\chi$ and
$\delta\psi_\mu$ on a set of Killing spinors, $\epsilon$, such that
$P\epsilon = 0$.  For a fivebrane with transverse directions $x^1$ and $x^2$,
the $\half$-SUSY projection takes the form
\eqn\halfsusy{P = \half(1+\gamma^{\overline1\overline2}T^{12}).}
Here $\gamma^{\overline1}$ and $\gamma^{\overline2}$ denote $\gamma$-matrices
with tangent space indices.  From the form of this projection, it is clear
that we have related the rotational $SO(2)$ symmetry in the $x^1$-$x^2$
plane with a $SO(2)$ subgroup of the $SO(3)$ automorphism group.  This is a
general property of $D-3$ branes constructed with scalars varying over a
two-dimensional base.
In fact, the projection \halfsusy\ and its resulting supersymmetry properties
is the basis for constructing the $SL(2,\ZZ)$ solution of $F$-theory \gsvy.
In that case the situation is quite clear, as parallel seven-branes (all
satisfying the identical projection $P$) may be combined to give an
elliptically fibered $K3$ surface.

In the $SL(3)/SO(3)$ case, however, we see that each fivebrane picks out
a specific embedding of $SO(2)$ within $SO(3)$, with considerable freedom on
how this embedding may be done.  Thus, making full use of $SO(3)$
invariance, the $\half$-SUSY projection generalizes to
\eqn\halfgen{P=\half(1+\gamma^{\overline1\overline2} \Lambda_{1a}\Lambda_{2b}
T^{ab}),}
where $\Lambda_{ab}(x)$ is a $SO(3)$ rotation matrix, allowing for different
fivebrane orientations at different points on the base.

While a single fivebrane preserves half of the supersymmetries, it is
well known that solutions with less supersymmetries may be constructed by
overlapping multiple branes.  From a supersymmetry point of view, this
corresponds to finding a set of commuting projections, $P_{(i)}$, each of
the form \halfgen, so that the only remaining supersymmetries are those
preserved by the complete set of $P_{(i)}$.  For the case at hand, it is easy
to see that, with a two dimensional base, it is impossible to construct
solutions based on more than one projection.  Equivalently, this indicates
that supersymmetric parallel fivebranes must have identical $SO(2)$
orientations within $SO(3)$, resulting in a solution preserving
exactly half of the supersymmetries.

The full $U$-symmetry of the fivebrane construction is brought out only when
the base is enlarged to three dimensions.  In this case, we may overlap two
fivebranes, constructed with {\it e.g.}\ the individual (commuting) projections
\eqn\twofive{\eqalign{
P_{(1)}&=\half(1+\gamma^{\overline1\overline2}T^{12})\cr
P_{(2)}&=\half(1+\gamma^{\overline2\overline3}T^{23}).}}
Note that we have made a rigid identification between the two $SO(3)$'s in
this case; as we see later, this cannot be the most general solution, but
nevertheless serves as a useful example.  While a third projection, $P_{(3)}
=\half(1+\gamma^{\overline1\overline3}T^{13})$, may be constructed, its
addition does not kill any more supersymmetries.  This is easily seen since
$P_{(3)}$ may be expressed as $P_{(3)} = P_{(1)} P_{(2)} +(1- P_{(1)})
(1-P_{(2)})$, and hence gives no further content than the combination of
$P_{(1)}$ and $P_{(2)}$ alone.

As an example of the above construction, we now turn to an explicit
parametrization of the $SL(3)/SO(3)$ vielbein $V_{ai}$ in terms of five
scalars.  Using $SO(3)$ invariance, we may write $V$ in the upper triangular
form
\eqn\vexpl{V=e^{\Phi_1/\sqrt{3}}\left[
\matrix{1&a&b\cr
0&e^{-(\sqrt{3}\Phi_1-\Phi_2)/2}&ce^{-(\sqrt{3}\Phi_1-\Phi_2)/2}\cr
0&0&e^{-(\sqrt{3}\Phi_1+\Phi_2)/2}}\right],}
where $\Phi_1$ and $\Phi_2$ are the two dilatonic scalars.  This
parametrization is motivated by the structure of the $SL(3)$ generators, in
particular with the dilatons corresponding to the Cartan generators
$\lambda^3$ and $\lambda^8$.

At this stage it is instructive to see the explicit form of the Lagrangian.
Defining the $SL(3)$ roots
\eqn\alpharoot{\eqalign{\alpha_1&={\sqrt{3}\over2}\Phi_1-\half\Phi_2\cr
\alpha_2&=\Phi_2,}}
we find
\eqn\lexpl{\eqalign{{\cal L}={1\over2\kappa^2}\sqrt{-g}
[R&-\third(\partial\alpha_1)^2-\third(\partial\alpha_2)^2
-\third(\partial\alpha_{12})^2\cr
&-\half e^{2\alpha_1}(\partial a)^2
-\half e^{2\alpha_2}(\partial c)^2
-\half e^{2\alpha_{12}}(\partial b-c\partial a)^2],}}
where $\alpha_{12} \equiv \alpha_1 + \alpha_2$ was introduced for
convenience and to highlight the nature of the $SL(3)$ symmetry.  From here
we see that only scalars $a$ and $b$ possess translational invariance, as
expected.

In contrast to the parallel fivebrane solution of \gsvy, here it is
not possible to separate the behavior of the gravity fields from that of the
scalars.  One may anticipate that this is the case since the external metric
of a $D-3$ brane contains a deficit angle.  Hence when several branes are
overlapped they would necessarily be affected by the presence of the others
which share only part of the transverse directions.  A natural choice for
the metric on the base is given by
\eqn\bmetr{ds^2=e^{2\phi_1(x)}dx_1^2+ e^{2\phi_2(x)}dx_2^2+
e^{2\phi_3(x)}dx_3^2.}
We note that making this choice has essentially forced us to consider only
``rigid'' fivebranes --- those with a globally fixed orientation between the
two $SO(3)$'s.

Demanding that the supersymmetry variations \susy\ vanish for Killing spinors
satisfying \twofive, we end up with a set of first order equations taking
the form
\eqn\ansatz{\eqalign{
\partial_1a&=-e^{\phi_1-\phi_2-\phi_3}\partial_2e^{\phi_3-\alpha_1}\cr
\partial_2a&=e^{-\phi_1+\phi_2+\phi_3}\partial_1e^{-\phi_3-\alpha_1}\cr
\noalign{\smallskip}
\partial_1b-c\partial_1a&=-e^{\phi_1+\phi_2-\phi_3}
\partial_3e^{-\phi_2-\alpha_{12}}\cr
\partial_3b\hphantom{-c\partial_3a\ }&=e^{-\phi_1+\phi_2+\phi_3}
\partial_1e^{-\phi_2-\alpha_{12}}\cr
\noalign{\smallskip}
\partial_2c&=-e^{\phi_1+\phi_2-\phi_3}\partial_3e^{-\phi_1-\alpha_2}\cr
\partial_3c&=e^{-\phi_1-\phi_2+\phi_3}\partial_2e^{\phi_1-\alpha_2},}}
in addition to the conditions
\eqn\morea{\partial_3a=\partial_2b-c\partial_2a=\partial_1c=0,}
and
\eqn\moreb{\eqalign{
\partial_i[\phi_1+{\textstyle{2\over3}}\alpha_1+\third\alpha_2]&=0
    \qquad i=2,3\cr
\partial_1[\phi_2+\third\alpha_1-\third\alpha_2]&=0\cr
\partial_3[\phi_2-\third\alpha_1+\third\alpha_2]&=0\cr
\partial_i[\phi_3+\third\alpha_1+{\textstyle{2\over3}}\alpha_2]&=0
    \qquad i=1,2.}}
While superficially these equations may appear quite formidable, they
actually have a very simple structure dictated by $SL(3)/SO(3)$ symmetry
considerations.  In \ansatz\ the three sets of equations, for non-dilatonic
scalars $a$, $b$, and $c$ respectively, are essentially generalized
Cauchy-Riemann-like equations, corresponding to individual fivebranes built
out of the pairs of fields $(a,e^{-\alpha_1})$, $(b,e^{-\alpha_{12}})$, and
$(c,e^{-\alpha_2})$.  Of course $\alpha_{12}$ is not independent, so
the equations do not separate, but are quite subtlely intertwined.  This is
simply a result of $SL(3)$ not being large enough to allow two independent
$SL(2)$ fivebranes.  Alternatively we note that there are two possibilities
for preserving a quarter of the supersymmetries, these being $K3\times K3$ or
$CY_3$, corresponding to $SL(2)\times SL(2)$ or $SL(3)$ brane configurations
respectively.

Before turning to overlapping brane solutions, we note that the ansatz gives
three $SL(2)/SO(2)$ special cases, obtained by setting either $b=c=0$, $a=b=0$
or $a=c=0$.  For example, in the first case, we would find
the Cauchy-Riemann equation
\eqn\sltsoln{\partial_ia(x_1,x_2) =-\epsilon_{ij}\partial_j
e^{-\alpha_1(x_1,x_2)},}
which is solved by complex analytic functions $\tau(z)$ where $\tau = a + i
e^{-\alpha_1}$ and $z=x_1 + i x_2$.  Thus in this case we have essentially
reproduced the $SL(2)$ solution of \gsvy.  One key difference, though, is
that by picking a ``rigid'' fivebrane orientation, the present ansatz gives
rise to the non-modular invariant relation between the metric and scalar
fields, $\phi_1 = \phi_2 = -\alpha_1/2 = \alpha_2$ (or $\phi = \log \tau_2$
in the notation of \gsvy), indicating that the global properties are not
fully addressed in this ``rigid'' ansatz.  In principle this issue is solved
by picking a more flexible ansatz allowing more freedom in the fivebrane
orientation.  It is this point that allows considerably more freedom in the
solutions, giving rise to a much richer structure of $T^3$ fibered $CY_3$'s
than the corresponding case of K3.

As anticipated, we observe that the metric fields $\phi_\mu$ do not separate
from the dilatons in the fivebrane ansatz.  Using the relation between
metric fields and dilatons, \moreb, as a hint, we may further restrict the
solution by introducing the three quantities
\eqn\phiabc{\phi_a(x_1,x_2)\qquad\phi_b(x_1,x_3)\qquad\phi_c(x_2,x_3),}
so that the metric ansatz now takes the new form
\eqn\newmeta{\phi_1 = -\half(\phi_a+\phi_b)
\qquad
\phi_2 = -\half(\phi_a+\phi_c)
\qquad
\phi_3 = -\half(\phi_b+\phi_c).}
To remain consistent with \moreb, the dilatons then must have the form
\eqn\newdila{\eqalign{\alpha_1 &= \half(2\phi_a+\phi_b-\phi_c)\cr
\alpha_2&=\half(-\phi_a+\phi_b+2\phi_c).}}
The beauty behind this choice of fields is that the Cauchy-Riemann-like
equations, \ansatz, now take the simplified form
\eqn\newansatz{\eqalign{
\partial_1a&=-e^{\phi_c-\phi_b}\partial_2e^{-\phi_a}\cr
\partial_2a&=\hphantom{-e^{\phi_c-\phi_b}}\partial_1e^{-\phi_a}\cr
\noalign{\smallskip}
\partial_1b - c \partial_1 a &= - e^{-\phi_a}\partial_3e^{-\phi_b}\cr
\partial_3b \hphantom{- c \partial_1 a\ } &=
    \hphantom{-} e^{-\phi_c}\partial_1e^{-\phi_b}\cr
\noalign{\smallskip}
\partial_2c&=\hphantom{e^{\phi_1}\>}-\partial_3e^{-\phi_c}\cr
\partial_3c&=e^{\phi_a-\phi_b}\partial_2e^{-\phi_c},}}
indicating the connection between $a,b,c$ and $\phi_a,\phi_b, \phi_c$
respectively.  This clearly shows the relation between the individual
fivebranes, their transverse directions, and its effect on the metric fields
through \newmeta.  For example, the fivebrane corresponding to $(a,\phi_a)$
is transverse in $x_1$ and $x_2$, and does not affect the metric component
$\phi_3$ in the longitudinal $x_3$ direction.  Viewed in terms of a
three-dimensional base, this picture is one of overlapping fivebranes living
on one-dimensional lines on the base.

In order to consider the overlapping solution of two fivebranes, we set both
$c=0$ and $\phi_c=0$.  As a result, \newansatz\ gives rise to
the second-order equations
\eqn\oversol{\eqalign{
(\partial_1^2+e^{-\phi_b}\partial_2^2)e^{-\phi_a}&=0\cr
(\partial_1^2+e^{-\phi_a}\partial_3^2)e^{-\phi_b}&=0,}}
which is an overlapping brane solution with $x_1$ being the common transverse
direction in the solution.  However, in addition to these transverse
laplacians, we also have the consistency requirement that
$\partial_3e^{-\phi_b}\partial_2e^{-\phi_a}=0$, so that either
$\phi_a$ or $\phi_b$ must be a function of $x_1$ only.  Taking the latter
case, the resulting fivebrane solution is described by the functions
$\phi_a(x_1, x_2)$ and $\phi_b(x_1)$, and corresponds to an overlapping
fivebrane and smeared fivebrane solution.

While it appears that the ``rigid'' ansatz only leads to a solution
with smeared out branes, it is anticipated that a more general $SO(3)$
ansatz would allow true overlapped fivebrane solutions.  On the other hand,
from a more general point of view, the solution may be viewed as a mapping
of the three-dimensional base into the five-dimensional moduli space.  While
this space is locally $SL(3)/SO(3)$, it is in fact an orbifold since points
must be identified under action of the $SL(3,\ZZ)$ $U$-duality.  It is thus
the pullback of the orbifold singularities that may be related to the
fivebrane configuration on the base.  In particular, singular lines, with
codimension two on the base, are then identified with the $SL(3)$ fivebranes.

While the above discussion has focused on $SL(3)$ solutions, we note that
cosmic-string-type solutions can be easily generalized to other extended
supersymmetric theories despite the fact that writing down explicit
first-order equations for the scalars generalizing \ansatz, \morea\ and
\moreb\ may seem terribly involved.  Let us recall the decomposition of
the solvable Lie algebra for $U/H$ ($Solv(U/H) \sim T{\cal M}_{scalars}$):
\eqn\solvd{Solv(U/H) = {\cal H} \oplus \Phi^+(U),}
where ${\cal H}$ is the Cartan piece, while $\Phi^+(U)$ is the positive
part of the root space of $U$. We have seen that the ``dilatonic''
(exponentiated) scalars in the Cartan subalgebra in turn appear in
combinations corresponding to the positive roots of the duality group.
Bearing in mind that
only the primitive roots of the duality group correspond to independent
``dilatons'', we now see that in each case the
number of pairs $(a_i, e^{\alpha_i})$ used to build individual solutions
is equal to the number of positive roots of the duality group. The
nested structure of the further decomposition of \solvd\  allows for more
simplifications in analyzing the solutions. For example, for the $D=8$
$SL(3) \times SL(2)$ case, the Cartan piece is three-dimensional, and
we find that $\Phi^+(U) = \Phi^+(E_2) \oplus {\cal D}^+_2$ where $\Phi^+(E_2)$
is the one-dimensional root space of the nine-dimensional $U$-duality group
(corresponding to the $RR$ scalar in type $IIB$), and ${\cal D}^+_2$ is the
weight space of the $SL(3) \times SL(2)$ irreducible representation to which
the nine-dimensional vectors are assigned (corresponding to
the three-dimensional abelian ideal, or in other words, the scalars with
translational symmetries).

\newsec{$T^3$ fibrations and Calabi-Yau manifolds}

In the previous section we have constructed a class of overlapping
fivebrane solutions with varying $SL(3)/SO(3)$ scalars.  We now show that
these solutions have a natural interpretation in terms of a $T^3$ fibered
Calabi-Yau manifold.  In particular, this fivebrane solution provides a
simple system where such $T^3$ fibrations may be studied in detail.

We begin by recalling that the $SL(2)$ $F$-theory solution may be described in
terms of a $K3$ fibration where a $T^2$ of constant volume but varying shape
is fibered over a $S^2$ base.  In particular, this solution is given in terms
of a function $\tau(z)$ that maps $z$, the complex coordinate on the sphere,
to $\tau$, the modular parameter of $T^2$.  Locally, any analytic map
$\tau(z)$ solves the equations of motion and preserves exactly half of the
supersymmetry.  However it is the global properties that give rise to the
intricacies of the solution.  In particular, exactly 24 strings are required
to give $c_1(m(S^2))=2$, leading to a fibered $K3$ surface (where $m$
defines a map from the base $S^2$ into the moduli space).

In the present case, since $SL(3)/SO(3)$ is locally the moduli space for
$T^3$ at constant volume, there is a similar picture of the fivebrane
solution in terms of a $T^3$ fibration.  As we have shown in the previous
section, an overlapping fivebrane solution preserving one quarter of the
supersymmetry naturally lives on a three-dimensional base (which is expected
to be topologically $S^3$ \gw).  Since the mapping is from an odd-dimensional
base to a five-dimensional moduli space, unlike the $SL(2)$ case, there are no
natural set of complex coordinates to work with.  On the other hand, for the
complete space to correspond to a Calabi-Yau 3-fold, there must exist a
choice of complex structure relating pairs of real coordinates.  While the
fivebrane ansatz, \ansatz, appears to give three sets of Cauchy-Riemann
conditions, suggesting an intertwined set of complex coordinates on the base
($x_1+i x_2$, $x_2 + i x_3$ and $x_3 + i x_1$), we find that this is in
fact not a natural choice.  Instead, the choice that is consistent with a
$T^3$ fibration is to pair each of the real coordinates on the base with a
corresponding coordinate on the internal $T^3$.  We now demonstrate this in
some detail.

Working with real geometry, we denote the coordinates on the base as $x^\mu$,
$\mu=1,2,3$.  Since the internal space is $T^3$, we introduce a set of
three periodic internal coordinates, $\xi^i = \xi^i + 1$ with $i=1,2,3$.
Combining the metric \bmetr\ on the base with the $SL(3)$ invariant
form of the metric on $T^3$, a natural choice for the six-dimensional metric
is simply
\eqn\cymet{ds^2 = e^{2\phi_\mu(x)}(dx^\mu)^2 + d\xi^i M_{ij}(x) d\xi^j,}
and defines a $T^3$ fibration over $S^3$.  Note that this metric is
block diagonal between the base and internal space, so that it
describes $T^3$ fibers that are always perpendicular to the base.

In order to better understand the $T^3$ fibration, it is essential to show
that the manifold defined in this manner is in fact complex.  As mentioned
above, we seek a complex structure relating internal and base coordinates in
pairs, so that at least locally the line element becomes
\eqn\cmplxmet{ds^2 = dz^\mu d\overline{z}^\mu.}
Comparing this with \cymet, and using the definition $M=V^TV$, we see that
natural complex coordinates are then of the form
\eqn\cmplxcoor{z^\mu = e^{\phi_\mu}x^\mu + i \delta_{\mu a} V_{a i} \xi^i,}
corresponding to a complex structure given by
\eqn\cyj{J^M{}_N=\left[\matrix{0&-e^{-\phi_{\mu}}
\delta_{\mu a}^{\vphantom{-1}}V_{a i}^{\vphantom{-1}}\cr
V^{-1}_{i b}\delta_{b\nu}^{\vphantom{-1}}e^{\phi_\nu}&0}\right].}
Note that this choice of complex structure involves some arbitrariness in
pairing up the coset and base $SO(3)$ symmetries; the particular form was
chosen above to agree with the fivebrane ansatz \ansatz.

Since $J$ is a function of $x_\mu$ and hence varies over the base, it is
important to check that it is actually integrable.  This check is easily
performed by examining the Nijenhuis tensor corresponding to $J$.  Since the
Nijenhuis tensor is constructed in terms of first derivatives of $J$, we
find a set of first order equations as an integrability condition on the
complex structure.  Remarkably these conditions are a subset of \ansatz, so
that in fact the complex structure is integrable for the fivebranes
constructed in the previous section.

Using the complex structure \cyj, we are now able to connect the real
parametrization of the fibered space with complex geometry.  In particular,
the K\"ahler form,
\eqn\kahler{k_{MN} \equiv g_{MP}J^P{}_N
= \left[\matrix{0&-e^{\phi_\mu}\delta_{\mu a}^{\vphantom{T}}
V_{ai}^{\vphantom{T}}\cr
V^T_{ja}\delta_{\nu a}^{\vphantom{T}}e^{\phi_\nu}&0}\right],}
may be determined to be covariantly constant for fivebranes solving \ansatz.
In fact, verification that $\nabla_M k_{NP}=0$ requires full use of all 15
conditions of the supersymmetric ansatz, \ansatz, \morea\ and \moreb.

Finally, to show that the fibered space is a Calabi-Yau 3-fold (at least
locally), we need to verify that \cymet\ is not only K\"ahler, but also Ricci
flat.  Since the Ricci tensor involves second derivatives, the calculation
is somewhat more tedious.  Nevertheless, explicit verification shows that
the overlapping fivebrane solution indeed gives a Ricci flat K\"ahler metric
\cymet.

So far we have only considered the local properties of this fibered space.
The global properties will clearly be related to the mapping of the  $S^3$
base into the 5-dimensional moduli space of $T^3$, the latter being the
orbifold $SL(3,\ZZ)\backslash SL(3,\IR)/SO(3)$.  While some of the periods are
obvious
({\it e.g.} the 1-cycles $a\to a+1$, $b\to b+1$ and $c\to c+1; b\to b+a$)
the overall picture is not so straightforward.  Nevertheless, we expect the
$T^3$ fibers to degenerate on singular {\it lines} on $S^3$, corresponding
to the loci of the fivebranes.  These global issues are currently under
investigation%
\foot{There seems to be an appropriate generalization of the usual
$SL(2;\ZZ)$ fundamental domain; in this $SL(3,\ZZ)$ case it would be
given by $-1/2 \leq a, b, c, \leq 1/2$ and $|{\vec \lambda}_2|  
\geq |{\vec \lambda_1}| \geq 1,$ where the period vectors are given by 
${\vec \lambda_1} = (a,\exp(-\alpha_1),0)$ and 
${\vec \lambda_2} = (b,c\exp(-\alpha_1),\exp(-\alpha_{12}))$.  It is the
boundaries of this domain and their intersections that give rise to
various fixed lines in the moduli space.}.

Writing down the metric \cymet\ and complex structure \cyj\ in fact provides
an explicit construction of a $T^3$ fibration of $CY_3$ \refs{\bbs, \syz, \gw}.
We recall that, for such a fibration, the $T^3$ must in fact be a special
Lagrangian submanifold, which essentially means that the pullback of the
K\"ahler form must vanish, and the pullback of the holomorphic Calabi-Yau
form must give the real volume form on $T^3$.  For the fivebrane
solution, the first condition is trivially satisfied since $k_{ij}$,
the restriction of the K\"ahler form \kahler\ to $T^3$, is automatically
vanishing.  For the Calabi-Yau form, we appeal to \cmplxcoor\ to write
\eqn\cyform{\Omega = i\prod_\mu (e^{\phi_\mu} dx^\mu
+ i \delta_{\mu a}V_{ai} d\xi^i),}
which is by construction holomorphic with respect to the complex structure
\cyj, and squares to the six-dimensional volume form.  One may also verify
that $\Omega$ is closed, provided the complex structure is integrable.  When
restricted to the $T^3$ fiber, the Calabi-Yau form simply becomes
$\Omega_{ijk}=\epsilon_{abc}V_{ai}V_{bj}V_{ck}=\det V \epsilon_{ijk}$, which
is indeed the expected result (since $\det V=1$).

In retrospect, this correspondence is not surprising at all, since the
fivebrane solution was explicitly constructed based on supersymmetry
requirements of the eight-dimensional theory.  Viewed from a $M$-theory point
of view, the $SL(3,\ZZ)$ part of the $U$-duality group is exactly the symmetry
of the eleven dimensional theory compactified to eight dimensions on $T^3$.
Thus while in the type $IIB$ picture the $T^3$ is not physical, in $M$-theory
it is certainly present.  Constructing a $SL(3)$ solution preserving a
quarter of the supersymmetries is equivalent in the $M$-theory picture to a
compactification to five-dimensions on a Calabi-Yau three-fold.  Of course
it is important to realize that the complete $U$-manifold involves not just
the $SL(3,\ZZ)$, but also the $SL(2,\ZZ)$ part of the $U$-duality group, where
the $SL(2,\ZZ)$ is rather obscure in the M-theory language.

\medskip
\noindent{\it Mirror symmetry as $T$-duality}

An implicit formulation of the mirror space to a Calabi-Yau manifold
was found in \syz, where it was argued that every Calabi-Yau that has
a mirror necessarily has a supersymmetric $T^3$ fibration.
Moreover, the mirror symmetry turns out to be just $T$-duality
on the $T^3$ fibers. Thus all the $U$-manifolds appearing here
should have mirrors, as they are constructed explicitly as $T^3$
fibrations.

We already saw that supersymmetry ``knows'' about the fibered structure
on the $U$-manifold.  As it turns out, it also ``knows'' about the mirror
$U$-manifold: $T$-duality on the fibers is nothing but a discrete
symmetry of the coset representative
\eqn\tdual{T:\> V \rightarrow (V^{-1})^T,}
corresponding to $T$-duality on all three abelian cycles of $T^3$.
The way the fibration metric is written in \cymet\ makes $T$-duality on
the fibers particularly simple. Since the metric is of diagonal form, the
duality on the internal part does not generate any torsion, and hence
its action is given by $M \rightarrow M^{-1}$.  This is guaranteed to be the
correct $T$-duality from the $M$-theory point of view since in that case $M$
{\it is} exactly the internal Kaluza-Klein metric. From a string theory point
of view, this inversion precisely corresponds to the interchange of
winding and momenta modes for all three compact dimensions%
\foot{More precisely, the string interpretation is obtained only upon
further compactification of $M$-theory on $S^1$.}.
Note, however, that while \tdual\ is contained in the $O(3,3;\ZZ)$ $T$-duality
group appropriate to a string compactification, it is not part of $SL(3,\ZZ)$.
Thus the inversion of the vielbein is a unique discrete symmetry of the
scalar manifold ${\cal M}$, and lies outside of the $U$-duality group%
\foot{A simple way to see this is to note that if the scalar matrix $M$
and its inverse were related under $SL(3,\ZZ)$, this relation would have the
form $M^{-1} = \Omega^T M \Omega$ where $\Omega$ is a $SL(3,\ZZ)$ matrix.
Since $\Omega$ has integer coefficients, it cannot vary continuously as the
five scalars are varied.  Thus every entry in $M^{-1}$ must be able to be
written as some linear combination of the terms already present in $M$.
However a simple calculation of the inverse shows that this is not in
general possible.}.

It is interesting to study the behavior of the fibration under
$T$-duality.  Suppressing all indices, we may rewrite the complex
structure \cyj\ and K\"ahler form \kahler\ in the block form
\eqn\origjk{
J = \left[\matrix{0 & -j^{-1}\cr j & 0}\right]
\qquad
k = \left[\matrix{0 & -\kappa^T\cr \kappa & 0}\right],}
where $j=V^{-1} \cdot e^{\phi}$ and $\kappa= V^T \cdot e^{\phi}$.
Here $e^\phi$ denotes the diagonal matrix of metric factors on the base,
$e^\phi = diag (e^{\phi_1}, e^{\phi_2}, e^{\phi_3})$.  In this form, it
immediately follows that the action of $T$-duality, \tdual, interchanges
$\kappa$ with $j$,
\eqn\coka{T:\> \kappa \leftrightarrow j,}
with corresponding interchange between $J$ and $k$.  As a result, this gives
an explicit realization of Calabi-Yau mirror symmetry, where deformations of
complex structure, $\delta j$, and K\"ahler class, $\delta \kappa$, are
interchanged between the original manifold and its mirror.  While this has
been a local statement, the picture nevertheless continues to hold in
general, since any globally well defined $\delta j$ on the $S^3$ base may
equally well apply to $\delta \kappa$ and vice versa.  Of course, the $T^3$
fibration structure is crucial, as it is the discrete freedom of choice
for the vielbein, \tdual, which corresponds to the choice between two mirror
$U$-manifolds.

\newsec{Comments and conclusions}

Starting from an attempt to construct $U$-branes transforming non-trivially
under $SL(3,\ZZ)$, we have ended up with an explicit {\it local}
description of $T^3$ fibered Calabi-Yau 3-folds.  The first order equations
resulting from the Killing spinor equations correspond in the Calabi-Yau
picture to the special lagrangian conditions.  By explicit construction, the
complex coordinates of the Calabi-Yau 3-fold are built from the coordinates
on the base each paired with a coordinate on the $T^3$ fiber.  This may be
contrasted with the $SL(2,\ZZ)$ ``stringy cosmic string'' construction of
\gsvy, where $K3$ is described in terms of complex tori fibered over
$CP^1$.  Since the two sets of complex coordinates used in \gsvy\ and
here are not related by an analytic map, our construction simply
corresponds to a different choice of the complex structure.
However (as pointed out in \syz) this description of $K3$ may be
related to the present case by a rotation of complex structure. In fact, it
is the (hyper-K\"ahler) $K3$ case that is special; general mirror pairs
of Calabi-Yau
$n$-folds presumably involve $T^n$ fibrations, giving a complex structure of
the form \cyj\ without any freedom of rotation of complex structure.

Another notable difference in the construction of an elliptically fibered
$K3$ from the one presented here is that, as mentioned before, the $K3$ case
only involves parallel branes.  One way to see this is that for the $SL(2)$
case there is a unique supersymmetry projection operator, the analogue of
\halfgen, which transforms as a singlet under the $SO(2)$ automorphism
group, resulting in a unique configuration preserving half of the
supersymmetries.  In the more general case, the freedom to pick different
$SO(n)$ orientations for the branes leads to many possibilities for
``closing" the manifold.  We observe that each additional overlapping brane
reduces the supersymmetry by a half, so that the seven dimensional
$SL(5,\ZZ)$ case preserves only a sixteenth of the supersymmetry, and
involves even fancier intersections of branes with different orientations%
\foot{In general, $SL(n)$ solutions are expected to preserve anywhere from
$1/2$ to $1/2^{n-1}$ supersymmetries, with the latter corresponding to true
$SL(n)$ solutions that do not factorize into separate pieces.  For example, 
with $SL(5)$ we would have a family of manifolds of the form
$K3 \times T^6$, $CY_3 \times T^4$, $CY_4 \times T^2$ and finally $CY_5$.}.
Once again this shows that $K3$ is a somewhat degenerate case, as it is
the only one where the $D-3$ branes are completely parallel.
It is of some interest to better understand the relation between the
distribution of branes on the base and the global properties of the
manifold such as the number of deformations of complex structure and
K\"ahler class.

One may speculate that, even with a complicated overlapping fivebrane
configuration, it would nevertheless be possible to study individual
branes within the solution, each given by a single ``dilaton-axion'' pair
$(a_i, e^{-\alpha_i})$ with a specific $SO(2)$ orientation on the base in
the notation of section 3.
Moreover, one would hope to evaluate the energy per volume for the individual
branes in this fashion and obtain a relation analogous to the
elliptically fibered case \grassi, where for the base $B$:
\eqn\ellipt{c_1(B) = -{1 \over 12} \sum N_a \delta_2 (D_a),}
which plays an important role in $F$-theory constructions. Physically
speaking, this is what one expects from the fivebranes on $S^3$. Indeed, the
effect of the branes on the curvature seems to be localized, but the
complicated global features such as the deficit angle of the solutions
do not allow an analysis of an isolated solution. Further exploration of
the properties of the moduli space should hopefully lead to progress in
this direction.

We reiterate that much remains to be done to achieve a better understanding
of the global features of the $U$-brane configurations.  Nevertheless,
the profound relation between supersymmetry and geometry as well as the
remarkable emerging picture of mirror symmetry makes this study worthwhile.

\vskip1truecm

\noindent
{\bf Acknowledgments:}
This work was supported in part by the U.~S.~Department of Energy under
grant no.~DOE-91ER40651-TASKB. RM is supported by a WL Fellowship.
JTL wishes to acknowledge the hospitality of the CERN theory division
during the visit where much of this work was conceived.

\appendix{A}{}

In this appendix we make an explicit connection between the $U$-scalars in
eight dimensions and the original type $IIB$ theory.  Starting in ten
dimensions, while the type $IIB$ theory contains a 4-form potential with
self-dual field strength, and hence does not admit a conventional Lagrangian
formulation, for our purposes the 4-form potential may be ignored, as it does
not give rise to any scalars in eight dimensions.  As a result, it is
sufficient to focus on the truncated type $IIB$ Lagrangian, written in natural
string coordinates as \bho
\eqn\iibact{\eqalign{
{\cal L}_{D=10} = \sqrt{-G^{(10)}}e^{-2\Phi^{(10)}}
[&R_{G^{(10)}}+4(\partial_M\Phi^{(10)})^2-{1\over2\cdot3!}(H_{MNP}^{(1)})^2\cr
& -e^{\Phi^{(10)}}({1\over2}(\partial_M\ell)^2+{1\over2\cdot3!}
(H_{MNP}^{(2)}-\ell H_{MNP}^{(1)})^2)],}}
with $H^{(i)} = dB^{(i)}$.  In this form, there is a clear division between
the $NSNS$ fields, $\{G^{(10)}, B^{(1)}, \Phi^{(10)}\}$, and the $RR$ fields
$\{B^{(2)}, \ell\}$ (ignoring the $RR$ 4-form potential).

For the type $IIB$ theory in ten dimensions, we may consider the fermions as
pairs of Majorana-Weyl spinors, transforming as a doublet under SO(2).  In
this case, introducing the Pauli matrices $\sigma^i$ acting on the SO(2)
index, the supersymmetries may be written as
\eqn\iibsusy{\eqalign{
\delta\Lambda&=\half[\Gamma^M\partial_M\Phi^{(10)}
-{\textstyle{1\over12}}H_{MNP}^{(1)}\Gamma^{MNP}\sigma^3]\eta\cr
&\qquad+\half e^{\Phi^{(10)}}[\Gamma^M\partial_M\ell i\sigma^2 +
{\textstyle{1\over12}}(H^{(2)}-\ell
H^{(1)})_{MNP}\Gamma^{MNP}\sigma^1]\eta\cr
\delta\Psi_M&=[\nabla-{\textstyle{1\over8}}H_{MNP}^{(1)}\Gamma^{NP}
\sigma^3]\eta\cr
&\qquad-{\textstyle{1\over8}}e^{\Phi^{(10)}}
[\Gamma^N\partial_N\ell i\sigma^2+{\textstyle{1\over6}}(H^{(2)}-\ell
H^{(1)})_{NPQ}\Gamma^{NPQ}\sigma^1]\Gamma_M\eta,}}
again with a clear split between the $NSNS$ and $RR$ fields.

Upon $T^2$ reduction to eight dimensions, both $H^{(1)}$ and $H^{(2)}$ give
rise to scalars.  Combined with $\Phi^{(10)}$, $\ell$ and the three scalars
parametrizing $T^2$, this gives a total of seven eight-dimensional scalars.
Since we are only interested in the scalar sector, dimensional reduction is
especially straightforward.  In order to end up in an eight-dimensional
Einstein frame, we take
\eqn\compactg{G_{MN}=\left[\matrix{e^{2\phi/3}g_{\mu\nu}\cr
&G_{ij}}\right],}
where the $T^2$ metric is
\eqn\ttwomet{G_{ij} = {e^{-\sigma}\over\tau_2}\left[\matrix{1&\tau_1\cr
\tau_1&|\tau|^2}\right],}
and the eight-dimensional dilaton is given by
\eqn\eightphi{\phi = \Phi^{(10)}+\half\sigma.}
Finally, defining $B_{ij}^{(k)}=b^{(k)}\epsilon_{ij}$, we end up with
\eqn\eightact{\eqalign{{\cal L}_{D=8} = \sqrt{-g}[R&
-\ha{|\partial_\mu\tau|^2\over\tau_2^2}
-{2\over3}(\partial_\mu\phi)^2-\ha(\partial_\mu\sigma)^2\cr
&-\ha e^{2\sigma}(\partial_\mu b^{(1)})^2
-\ha e^{2\phi-\sigma}(\partial_\mu\ell)^2
-\ha e^{2\phi+\sigma}(\partial_\mu b^{(2)}-\ell \partial_\mu b^{(1)})^2],}}
which may be compared with \lexpl.  As a result, we identify the $SL(3)$
scalars as
\eqn\slthreei{\{a,b,c\}=\{b^{(1)},b^{(2)},\ell\}\qquad
\{\alpha_1,\alpha_2\}=\{\sigma,\phi-\half\sigma\},}
and the $SL(2)$ scalars as simply the complex structure of the $T^2$, namely
$\{\tau_1,\tau_2\}$.

Working out the supersymmetry of \eightact\ is somewhat more involved.
Upon dimensional reduction, the eight-dimensional dilatino ($\lambda$)
is shifted according to
\eqn\dilshift{\lambda=e^{\phi/6}(\Lambda-\half\Gamma^i\Psi_i),}
where the exponential factor accounts for transforming from the string to
the Einstein frame.  This latter transformation also shifts the gravitino so
that
\eqn\graveight{\psi_\mu=e^{-\phi/6}\Psi_\mu
-{\textstyle{1\over3}}\gamma_\mu \lambda.}
The resulting eight-dimensional gravitino variation becomes
\eqn\egravdel{\eqalign{\delta\psi_\mu=&[\nabla_\mu
-{i\over4}{\partial_\mu\tau_1 \over\tau_2}\gamma^9]\epsilon\cr
&-{1\over4}[e^\sigma\partial_\mu
b^{(1)}(\gamma^9i\sigma^3)+e^{\phi-{1\over2}\sigma}\partial_\mu\ell(i\sigma^2)
+e^{\phi+{1\over2}\sigma}(\partial_\mu b^{(2)}-\ell\partial_\mu b^{(1)})
(\gamma^9i\sigma^1)]\epsilon,}}
where $\epsilon=e^{-\phi/6}\eta$ and $(\gamma^9)^2=1$ is the chirality
operator in eight dimensions
($\gamma^9\equiv i\gamma^0\gamma^1\cdots\gamma^7$).
In performing the reduction of \iibsusy, we have made use of the fact that
the ten-dimensional spinors have definite chirality, $\Gamma^{11}\eta=\eta$.
Comparing \egravdel\ to \susy\ (making use of the $SL(3)$ interpretation
\slthreei), we find the $SO(3)$ generators
\eqn\sothreegen{T^a=\{\sigma^2,-\gamma^9\sigma^1,\gamma^9\sigma^3\},}
resulting in
\eqn\gravdelf{\delta\psi_\mu=[\nabla_\mu-{i\over4}{\partial_\mu
\tau_1\over\tau_2}\gamma^9+{1\over4}Q_\mu^{ab}T^{ab}]\epsilon.}
For the remaining spin-$\half$ fields, the dilatino may be combined with the
``internal'' components of the gravitino, $\Psi_8$ and $\Psi_9$, in the
combination
\eqn\combi{\chi^a=\left\{\eqalign{
\sigma^2&(-\third\lambda+\half e^{\phi/6}\Gamma^i\Psi_i
+\third e^{\phi/6}(\Gamma^8\Psi_8-\Gamma^9\Psi_9))\cr
\gamma^9\sigma^1&(-\third \lambda-\half e^{\phi/6}\Gamma^i\Psi_i
+\third e^{\phi/6}(\Gamma^8\Psi_8-\Gamma^9\Psi_9))\cr
-\gamma^9\sigma^3&(\hphantom{-}{\textstyle{2\over3}}\lambda
\hphantom{+\half e^{\phi/6}\Gamma^i\Psi_i\ \,}
+\third e^{\phi/6}(\Gamma^8\Psi_8-\Gamma^9\Psi_9))}\right\},}
with resulting variation
\eqn\chidelf{\delta\chi^a=-\ha\gamma^\mu P_\mu^{ab}T^b\epsilon
-{1\over6}\gamma^\mu[{\partial_\mu\tau_2\over\tau_2}
-i{\partial_\mu\tau_1\over\tau_2}\gamma^9]T^a\epsilon.}
The supersymmetry variations \gravdelf\ and \chidelf\ indicate the form of
the additional $SL(2)$ contributions that were ignored in \susy.
The above dimensional reduction demonstrates the explicit correspondence
between the type $IIB$ fields and the $U$-scalars of the eight dimensional
theory.

Finally, we note that the $U$-duality group in $D$ dimensions has a convenient
type $IIB$-inspired decomposition \solvv\ that makes the original
ten-dimensional $SL(2)_U$ symmetry apparent:
\eqn\decom{E_{r+1} \rightarrow SL(2, \IR)_U \otimes GL(r, \IR),}
where $r$ stands for the number of compact dimensions. Note that only the
first factor in \decom\ mixes $RR$ and $NSNS$ states; $GL(r, \IR)$ is just the
isometry group of the classical moduli space of the $T^r$ torus.

\vfil\eject
\listrefs
\bye